\documentclass[review,3p,times,onecolumn]{elsarticle}
\usepackage{lineno,hyperref}
\modulolinenumbers[5]
\usepackage{xcolor}
\usepackage{comment}
\usepackage{latexsym}
\usepackage{epsfig}
\usepackage{amsmath}
\usepackage{amsfonts}
\usepackage{amssymb}
\usepackage{amsthm}
\usepackage{graphicx}
\usepackage{wrapfig}
\usepackage{pgf}
\usepackage{caption}
\usepackage{subcaption}
\usepackage{natbib}
\usepackage{tikz}
\usepackage[font=normalsize]{caption}

\journal{J.~Non-Newt.~Fluid Mech.}

\begin{document}	

	\begin{frontmatter}
	
	\medskip  
	
	\title{Statistics and spectral analysis of turbulent duct flows with flexible and rigid polymer solutions}
	
	\medskip
	
	\author[1]{Rodrigo~S.~Mitishita\corref{cor1}}
	\ead{rodrigo.seiji06@gmail.com} 
	\cortext[cor1]{Corresponding author}

	\author[1]{Gwynn~J.~Elfring} 
	
	\author[1,2]{Ian.~A.~Frigaard} 
	
	\address[1]{Department of Mechanical Engineering, University of British Columbia, 6250 Applied Science Ln, Vancouver, BC, V6T 1Z4, Canada}
	
	\address[2]{Department of Mathematics, University of British Columbia, 1984 Mathematics Rd, Vancouver, BC, V6T 1Z2, Canada}
	
	\begin{abstract}
		We present an experimental investigation of turbulent drag reduction with flexible and rigid polymer solutions. The flexible polymer is partially hydrolyzed polyacrylamide (HPAM) and the rigid polymer is xanthan gum (XG). The experiments are carried out at low drag reduction ($\%DR < 40$), high drag reduction ($\%DR > 40$) and maximum drag reduction (where the velocity profile $U^+$ roughly matches Virk's asymptote). We compare velocity profiles, streamwise and wall-normal Reynolds stresses and power spectra of streamwise velocity fluctuations measured by Laser Doppler Anemometry (LDA). Our results show that the effects of both XG and HPAM polymers on turbulence are similar, provided that the Reynolds numbers and $\%DR$ are also similar. At high levels of $\%DR$, the power spectral densities of streamwise velocity fluctuations of both XG and HPAM flows show a power-law decay near $-3$ instead of $-5/3$ in the inertial range. A slope of the power spectra of $-3$ was recently interpreted as evidence of elasto-inertial turbulence (EIT) in polymer jets. At relatively low concentrations, we observe that flexible polymer solutions are more effective at reducing drag, while XG only reaches MDR at very high concentrations. Thus, we hypothesize that the formation of polymer aggregates with higher concentrations contributes to increase viscoelasticity, and thus $\%DR$, with XG and other rigid polymer solutions.
	\end{abstract}
	
	
\end{frontmatter}

%%%%%%%%%%%%%%%%%%%%%%%%%%%%%%%%%%%%%%%%%%%%%%%%%%%%%%%  Main text %%%%%%%%%%%%%%%%%%%%%%%%%%%%%%%%%%%%%%%%%%%%
\section{Introduction} \label{S1}
 
The discovery of drag reduction in a turbulent flow by minute additions of polymers by \citet{toms1977} led to significant energy savings in the pumping of liquids across long distances, such as crude oil transport \citep{white2008}, fire-fighting operations \citep{soares2015} and HVAC systems \citep{gasljevic1997}. Even a few parts-per-million (ppm) addition may lead to a turbulent drag reduction (DR) of 70\%. This important finding paved the way to further experimental and numerical research efforts to understand the mechanisms through which polymer additives interact with a turbulent flow. Polymer solutions for drag reduction can be formulated with flexible polymers, which are usually synthetic such as polyethylene oxide (PEO) and polyacrylamide \citep{ptasinski2001,mohammadtabar2020}, or rigid molecules such as the biopolymers xanthan gum (XG) and diutan gum \citep{santos2020}. While both of these types of additives are able to provide significant reductions in turbulent drag, the proposed mechanisms for drag reduction appear to differ.

In general, flexible polymers are more desirable for DR applications, since a small concentration is enough to greatly reduce drag without a large increase in the viscosity of the fluid. Furthermore, constitutive models such as the Oldroyd-B and the finitely extensible nonlinear elastic dumbbell model with the Peterlin approximation (FENE-P) provide a framework for many numerical simulations of turbulent flows with flexible polymer solutions with accurate comparisons to experiments \citep{ptasinski2001}. Rigid polymer solutions, in contrast, have received less attention. One of the key differences observed in experiments is that rigid polymer solutions require much larger concentrations than flexible polymers to reach high levels of DR \citep{escudier2009a,warwaruk2021}. Moreover, rigid polymeric fluids differ from flexible ones in terms of their resistance to mechanical scission (or degradation) in turbulent flow \citep{pereira2013, soares2015, soares2020}. In fact, degradation in rigid polymers appears to occur due to destruction of aggregates instead of molecular scission \citep{pereira2013}. Even though there are only a few direct comparisons of turbulent flows with rigid and flexible polymer solutions \citep{pereira2013, mohammadtabar2017, warwaruk2021}, turbulence statistics appear to be quite similar to each other \citep{escudier2009a, mohammadtabar2017, shaban2018, warwaruk2021}, both displaying a significant decrease in wall-normal and shear Reynolds stress profiles as DR increases.

Over the years, two main theories were proposed to explain DR. The elastic theory by \citet{tabor1986} states that elastic stresses from the stretched polymer molecules dampen small-scale turbulent fluctuations. Drag reduction is commonly associated with viscoelastic properties in polymer solutions. The ratio of elastic to viscous forces in a shear dominated flow is quantified by the Weissenberg number $Wi$, the product of the extensional relaxation time of the viscoelastic fluid and the characteristic shear rate of the flow \citep{white2008}. Experimentally, the extensional relaxation time can be estimated with Capillary Breakup Extensional Rheometer (CaBER) experiments. The extensional relaxation time correlates with turbulent drag reduction in turbulent pipe flows by numerous studies \citep{owolabi2017,shaban2018,mohammadtabar2020}, and is directly related to the flexibility of the polymer molecules \citep{white2008}. Conversely, dilute polymer solutions of rigid molecules such as XG show negligible extensional relaxation times \citep{jaafar2010, mohammadtabar2020}, but the storage ($G'$) and viscous moduli ($G''$) can be measured via small amplitude oscillatory shear (SAOS) experiments in semi-dilute solutions \citep{pereira2013}. A larger $Wi$, obtained by either increasing the Reynolds number $Re$ or polymer concentration, reduces drag up until the maximum drag reduction \citep{virk1970} (MDR) state, beyond which further increases in $Wi$ do not contribute to drag reduction. At MDR, the shear Reynolds stress profile approaches zero, indicating marginal levels of turbulence. Interestingly, flexible and rigid polymers are bound by the same MDR asymptote, an observation confirmed by experiments \citep{virk1997,escudier2009a,mohammadtabar2017,mitishita2022} and theoretical studies \citep{benzi2005}. 

The viscous theory by \citet{lumley1969} suggests turbulent structures are suppressed by the increase in extensional viscosity due to stretching of the polymer molecules, causing drag reduction. To support the viscous theory of drag reduction, a large number of theoretical analyses of turbulent flows have been put forward, mostly based on the large increase in effective viscosity observed in viscoelastic fluids submitted to extensional strain rates, due to polymer stretching \citep{benzi2006}. Theoretical studies have been performed with the FENE-P model to predict the TDR up to maximum drag reduction (MDR) \citep{benzi2006,benzi2010}. However, the viscous theory of DR does not exclude viscoelastic properties of polymer solutions, since most theoretical studies still use the FENE-P model, and strain hardening is a well known occurrence in viscoelastic fluids. Thus, available theories may not be capable of explaining DR by rigid polymer solutions \citep{soares2015, xi2019}. Furthermore, turbulent flows of purely viscous, shear-thinning fluids have been investigated, both numerically \citep{rudman2004,singh2017b} and experimentally \cite{mitishita2021}, resulting in a generally low percentage of TDR, which implies that large DR appear to be closely related to fluid elasticity.

Significant advance in uncovering the mechanisms of drag reduction with flexible polymers were due to high resolution direct numerical simulations (DNS) with the FENE-P model, which generally provided good agreement with experimental observations. Moreover, DNS data can yield information that is difficult to access via experiments, such as the three-dimensional velocity field at high temporal resolution, and also the polymer stress field \citep{shahmardi2019}, which is not possible to obtain experimentally. The dynamics of of coiling and stretching of polymer molecules, and their relation to TDR, was uncovered by DNS in \citet{xi2012a}, with energy transfers further investigated by \citet{pereira2017c}, where polymer molecules cycle between a coiled and stretched cycle, anticorrelated to DR. In other words, as coiled molecules begin to extend, velocity fluctuations and shear Reynolds stresses increase with drag (active state). The extending of the molecules then absorb energy from turbulent structures, reducing drag. The low turbulence state causes the polymer molecules to coil (hibernating state), and so re-starting the cycle. MDR was then found to be a dominant hibernating turbulence state.

Relating viscoelastic effects to DR, \citet{samanta2013} studied the phenomenon named elasto-inertial turbulence (EIT), where it was found that viscoelasticity may trigger turbulence in pipe flows of polymer solutions at much lower Reynolds numbers than what is typically observed in Newtonian flows. As $Re$ is increased, the transitional regime is ``skipped", and the flow reaches a turbulent state where the reported friction factor coincides with Virk's MDR. From then, numerous DNS studies with the FENE-P model described the turbulent structures of EIT, which resemble a 2-dimensional, marginally turbulent flow at low $Re$ \citep{dubief2013, sid2018}. Furthermore, \citet{choueiri2018} demonstrated that, at certain combinations of Reynolds and Weissenberg numbers, a turbulent, drag-reducing flow of a dilute polymer solution may become laminar once the polymer concentration is increased up to a certain value. If the amount of polymer is increased further, the laminar state becomes unstable and MDR is reached. They concluded that EIT does not appear to be related to Newtonian turbulence, as the flow structures are considerably different.

In addition to turbulence statistics obtained from DNS and experimental measurements, useful information can be acquired from the power spectral densities of velocity fluctuations. It is especially useful when comparing numerical simulations to experiments, since experimentalists usually have access to power spectral densities from velocity signals acquired over time at a high data rate. In DR, a sharp drop in the power spectrum is observed for larger frequencies, meaning that the smaller length scale eddies lose energy, modifying the energy cascade in comparison to a Newtonian flow \citep{warholic1999,warholic1999influence,thais2013,vonlanthen2013,mitishita2022}, where a $-5/3$ slope is seen in the inertial range \citep{kolmogorov1941}. 

Recently, a universal power-law scaling of the power spectra of fluctuations of polymer concentration in submerged turbulent jets of $-3$ was observed \citep{yamani2021}, instead of a $-5/3$ slope. The $-3$ power law in DNS of EIT was also observed in simulations of isotropic viscoelastic turbulence, though only for moderate Deborah numbers \citep{valente2016}. The slope of $-3$ of the power spectra has also been observed in a turbulent duct flow of wormlike micellar gels, semi-dilute solutions of partially hydrolyzed polyacrylamide (HPAM, a flexible polymer) and xanthan gum (a rigid polymer) \citep{mitishita2022}. This finding could hint to similar turbulent structures in the drag-reducing flows of rigid, flexible polymers and also surfactant solutions. To the best of our knowledge, the power spectral densities of polymer solutions in turbulent flows at distinct levels of DR have only been investigated experimentally by \citet{warholic1999influence}, albeit with limited discussions related to the transition between each DR levels, and only with a flexible polymer type.

As a continuation of the experiments presented by our previous work \citep{mitishita2022}, and to further elucidate the similarities or differences between turbulent drag reduction with rigid and flexible polymers, we analyse the turbulence statistics and power spectral densities of velocity fluctuations of HPAM and XG at high Reynolds numbers via laser Doppler anemometry (LDA). We investigate three levels of drag reduction at similar Reynolds numbers: low drag reduction, ($LDR, DR < 40\%$), high drag reduction (HDR, $DR > 40\%$) and maximum drag reduction (MDR), where the percentage of DR is over 60\%. We carry out our experiments at high data acquisition frequencies of over 2000 $Hz$, which provides greater accuracy in the high-wavenumber range spectra \citep{adrian1986}. With the high-resolution turbulence statistics and power spectral densities, we attempt to identify characteristics of transition from LDR to HDR and MDR regimes with rigid and flexible polymer solutions. The paper is organized as follows: firstly, we outline the experimental setup and procedures. Next, we present the time-averaged flow statistics, followed by the power spectral densities of streamwise velocity fluctuations at three different positions in the duct. We follow the data presentation with a discussion and conclusions of the main findings.

\section{Experiments} \label{S2}

\subsection{Flow loop and LDA setup} \label{S2-1}

\begin{figure*}
	\centering
	\includegraphics*[width=160mm]{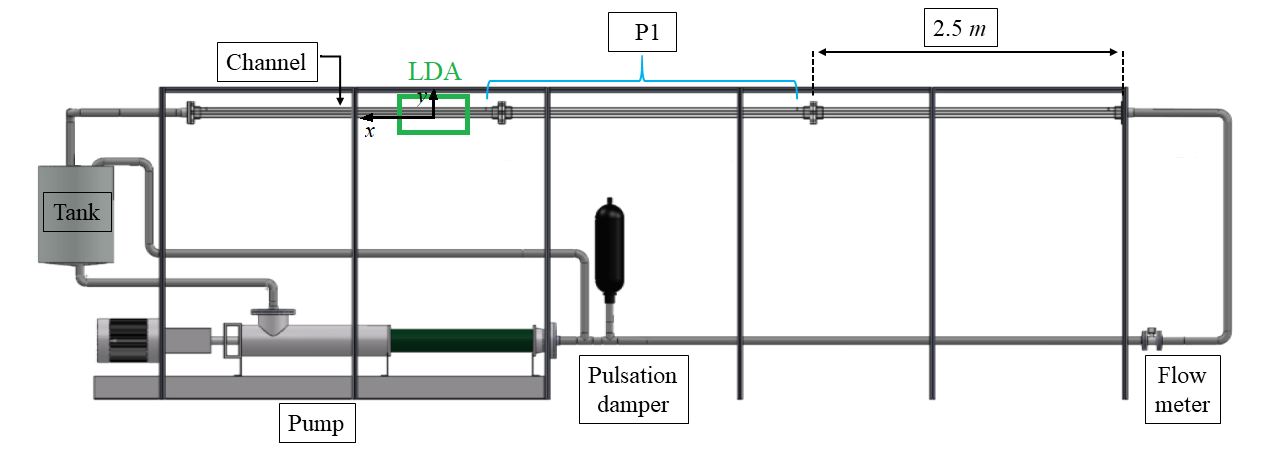}
	\caption{Flow loop schematic.}
	\label{fig1}
\end{figure*}

We only present a summary of our experimental setup, as more details can be found in our previous work \citep{mitishita2021}. We carry out our experiments in a horizontal, pump driven flow loop through a transparent $7.5 \, m$-long rectangular duct. A schematic of the flow loop is presented in \hyperref[fig1]{Figure~\ref*{fig1}}. The rectangular test section (``Duct'' in \hyperref[fig1]{Figure~\ref*{fig1}}) is made of three 2.5 $m$ long, clear acrylic ducts, with inner dimensions of width $W = 50.8$ \textit{mm} x height $H = 25.4$ \textit{mm} $= 2h$, where $h$ is the duct half-height, with hydraulic diameter $D_h = 2WH/(W+H)$ of 33.8 $mm$. The fluid in the storage tank is driven by a Netzsch NEMO progressing cavity pump, capable of a maximum flow rate of 20 $l/s$. A Parker pulsation damper was installed after the pump discharge to prevent pressure fluctuations. An Omega FMG 606 magnetic flow meter, with accuracy of $\pm0.5\%$ of full scale is used to measure the average flow rate, and the average bulk velocity $U_b$ can be calculated by dividing the flow rate by the cross sectional area of the duct. The pressure drop along 2.5 \textit{m} of the rectangular test section is measured by a PX419 pressure transducer with $\pm0.08\%$ accuracy relative to the full range of 0 to 15 \textit{psi} at a data acquisition rate of 3 \textit{hz}. The temperature of the fluid in the tank is measured by an Omega TC-NPT pipe thermocouple, of $\pm0.5\%$ accuracy. During a data collection procedure with the same fluid, the temperature increased by less than 2$^\circ$C, as we do not have a temperature controller. Remote control of the pump speed, as well as the signal acquisition from the thermocouple, pressure transducers and flow meter are provided by the National Instruments LabVIEW software and compact data acquisition modules. We perform laser Doppler velocimetry experiments over 5 \textit{m} (or approximately 150$D_h$) downstream of the test section inlet, where we consider the flow to be fully developed.

Instantaneous velocity data is obtained with a two-component Dantec Dynamics FlowExplorer LDA setup in back-scatter mode.  The laser supplies a pair of 532 \textit{nm} wavelength laser beams for velocity measurements in the streamwise direction (or \textit{x}-direction, for positive streamwise velocity $U$), and a pair of 561 \textit{nm} wavelength laser beams for velocity measurements in the wall normal direction (\textit{y}-direction, for positive wall normal velocity $V$). The optical setup allows for an ellipsoidal measurement volume of 0.1 \textit{mm} diameter and 0.3 \textit{mm} length with a 150 \textit{mm} focal length lens. A frequency shift of 80 \textit{Mhz} is applied to each laser by a Bragg cell. The probe is connected by to the Burst Spectrum Analyser (BSA) signal processor for data acquisition with the Dantec BSA software. Seeding particles manufactured by Dantec Dynamics, of 5 $\mu$\textit{m} average size, are used for the experiments. The uncertainty of the LDA system measurements, according to the factory calibration certificate, is 0.1\% of the mean with a 95\% confidence level. 

Since the LDA provides a point-wise velocity measurement, we employ a custom traverse system consisting of a linear motorized actuator by Zaber Motion Control of a total travel distance of 100 \textit{mm} with 25 $\mu$\textit{m} accuracy for vertical motion, to allow traverse in the \textit{y}-direction. Attached to the vertical traverse is another linear actuator by Zaber for horizontal movement in the spanwise direction. This actuator has a total travel distance of 250 \textit{mm} with 63 $\mu$\textit{m} accuracy. We traverse the probe in the spanwise direction only once to find the mid-point of the duct, where we perform all velocity measurements. The procedure of finding the position of the walls relative to the measurement volume is detailed in \citep{mitishita2021,mitishita2022}, considering that the refractive indices of our polymer solutions are the same as water.

\subsection{Fluid preparation and rheometry} \label{S2-2}

Our experiments consist of a comparison between turbulent flows of XG and HPAM at three levels of drag reduction. At low drag reduction, we investigate the flow of XG at 150ppm and HPAM at 7.5ppm. To probe the high drag reduction regime, we study the turbulent flow of XG and HPAM at 500ppm and 30ppm, respectively. Finally, the MDR regime is investigated with XG and HPAM at 1500ppm and 100ppm, respectively.

The mixing and experimental procedure are similar to our previous work \citep{mitishita2022}, as the same polymer powders of partially-hydrolyzed polyacrylamide (HPAM, Poly-plus RD by MI-Swaco) and xanthan gum (XG, Keltrol T by CP Kelco) are also employed here, albeit at different concentrations. We carefully pour the desired mass of each polymer in approximately $220 \, l$  of tap water in the flow loop tank. During this step, the water was recirculated at a high speed to form a vortex in the tank, ensuring mixing. Once all the polymer is introduced to the tank, we recirculate the solution for an additional 20 minutes at approximately 3 \textit{m/s}. The final, clear solutions are then rested overnight for homogenization. Samples for rheological assessment are collected after one complete data set (a full 26-point velocity profile measurement with LDA). At least three data sets with each fluid concentration are measured, each one with different average flow velocities. The bulk velocities of each data set, after which the samples were collected for rheometry are listed in \hyperref[table1]{Table~\ref*{table1}}.

The rheology of the polymer solutions is investigated with a high-resolution Malvern Kinexus Ultra+ rheometer. We use parallel plates with 40 \textit{mm} in diameter with a 0.3 \textit{mm} gap to allow for high shear rate measurements. The shear position in the parallel plates is at $r/R = 0.75$, where $R$ is the radius. The fluid temperature is controlled by a Peltier system with $\pm 0.1^{\circ}C$ accuracy, and it is left to stabilize in the geometry for 5 minutes before each test. Fluid characterization is performed via steady, shear-rate controlled flow curves. 

Considering that the fluid is well mixed before the turbulent flow experiments, and relatively small temperature changes do not affect the rheology of the fluid significantly, we assume that any significant variations in the flow curve after each flow loop data-set are consequences of degradation. Therefore, we average the steady flow curves from the three data sets for each polymer solution in \hyperref[fig2]{Figure~\ref*{fig2}}. Degradation does not appear to be significant overall, since most error bars (standard deviation from the average at each point) are quite small in most data points in the flow curve. Even in the dilute, flexible polymer solutions such as the 7.5 and 30 ppm HPAM, where degradation is generally more prevalent, according to previous experiments \citep{pereira2012}, we did not measure significant changes between each sample of the same fluid. The rigid XG solutions in particular are well known to be quite resistant to degradation, especially at higher concentrations \citep{soares2020}. The average viscosities of XG are fit to a Carreau-Yasuda (CY) constitutive equation: 

\begin{figure*}
	\centering
	\includegraphics*[width=160mm]{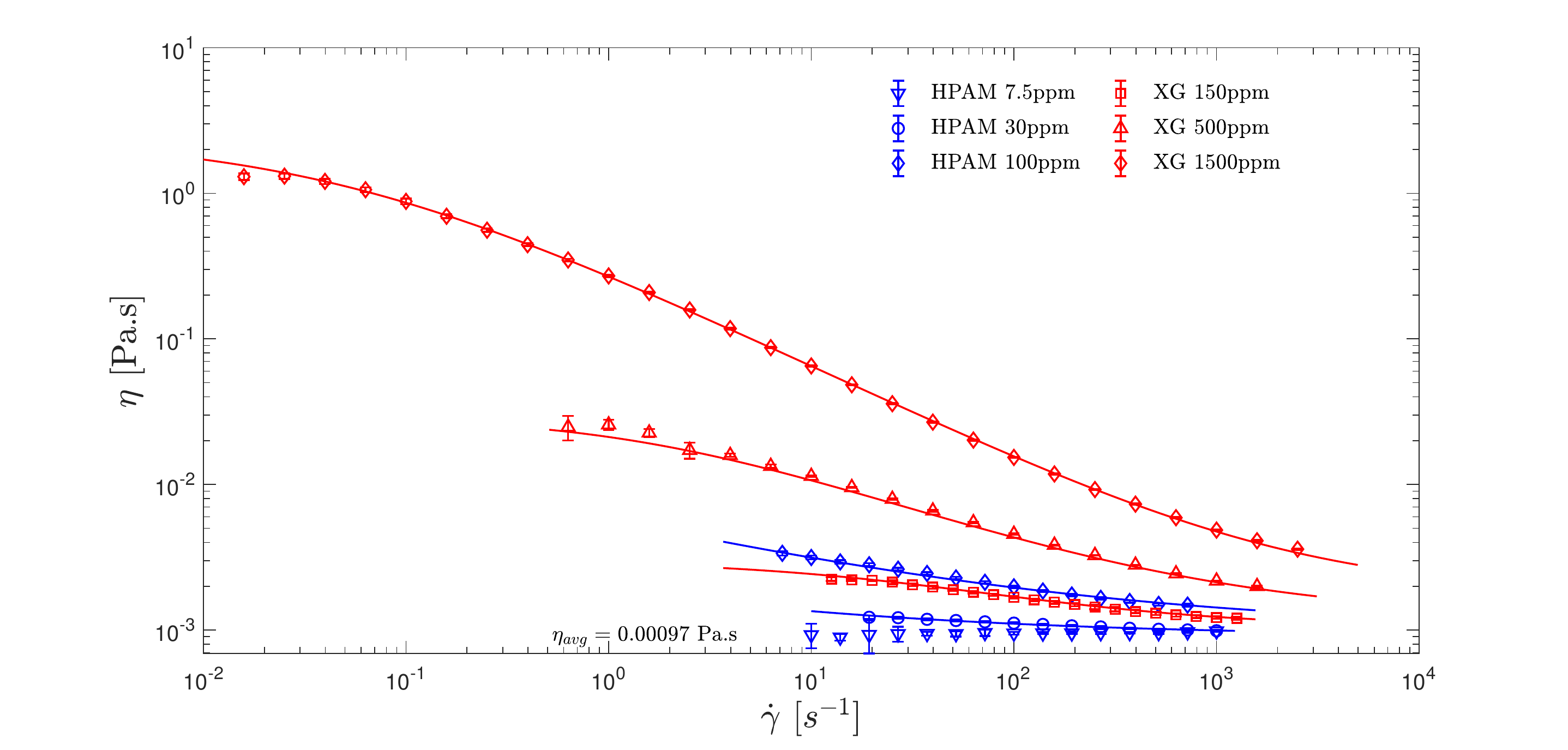}
	\caption{Steady flow curves of HPAM and XG solutions. Curve fits to the Carreau-Yasuda and Sisko models are shown as solid lines.}
	\label{fig2}
\end{figure*}

\begin{equation} \label{eq-CY}
	\eta = \eta_{\infty} + \frac{\eta_{0} - \eta_{\infty}}{(1 + (\lambda_{CY} \dot{\gamma})^a)^{n/a}},
\end{equation}

\noindent where $\eta_{0}$ is the zero-shear viscosity, $\eta_{\infty}$ is the infinite-shear-rate viscosity, $\lambda_{CY}$ is a constant with dimension of time, $n$ is a power-law index and $a$ is a fitting parameter \citep{yasuda1981}. Due to the difficulty in low shear-rate measurements with the low concentration 30 and 100ppm HPAM solutions, we fit the viscosity data to the Sisko constitutive equation \citep{sisko1958}:
\begin{equation} \label{eq-SK}
	\eta = \eta_{\infty} + K \dot{\gamma}^{n-1},
\end{equation}
where $K$ is the consistency index. All curve fits were performed via a nonlinear least squares procedure from MATLAB. The 7.5ppm HPAM solution showed constant viscosity measurements close to water, and thus we averaged the viscosity at all imposed shear rates as shown in \hyperref[fig2]{Figure~\ref*{fig2}}. 

In agreement with previous experiments \citep{escudier2009a,warwaruk2021}, a much larger concentration of xanthan gum is required to achieve the same levels of drag reduction of a flexible polymer solution such as HPAM (e.g. MDR can be achieved with 100ppm of HPAM, but 1500ppm of XG were necessary for MDR). The parameters of the curve-fitting procedures are found in \hyperref[table1]{Table~\ref*{table1}}.

\begin{table*}[t]
		\centering
		\begin{tabular}{ *{8}{c} }
			\hline
			Concentration & $U_b$ [$m/s$] & $\eta_{0}$ [$Pa.s$] & $K$ [$Pa.s^n$] & $\eta_{\infty}$ [$Pa.s$] &  $\lambda_{CY}$ [$s$] & $a$ [-]  & $n$ [-] \\
			\hline
			\multicolumn{8}{c}{HPAM} \\
			\hline	
			7.5 ppm & 1.86, 2.26, 2.76 & - & 0.00097 & - & - & - & 1 \\
			30 ppm & 1.84, 2.24, 2.74 & - & 0.00098 & 0.0009 & - & - & 0.66 \\
			100 ppm & 1.72, 2.22, 2.71 & - & 0.0048 & 0.001 & - & - & 0.65 \\
			\hline
			\multicolumn{8}{c}{XG} \\	
			\hline
			150 ppm & 1.86, 2.26, 2.76 & 0.003 & - & 0.001 & 0.036 & 0.709 & 0.619 \\
			500 ppm & 1.85, 2.25, 2.75 & 0.0295 & - & 0.0012 & 0.562 & 0.769 & 0.539 \\
			1500 ppm & 2.24, 2.74, 3.24, 3.71 & 2.436 & - & 0.0018 & 21.22 & 0.616 & 0.674 \\
			\hline
		\end{tabular} \caption{\label{table1} Fluid parameters for turbulent flow experiments. All fluids have density $\rho = 999~kg/m^3$. The $U_b$ column shows the average bulk velocity in each duct flow experiment after which the samples were collected for rheology assessment.}
\end{table*}

Finally, we note that SAOS tests at different frequencies were attempted with the dilute HPAM solutions, but inertial effects resulted in unsuccessful experiments, even at low frequencies. Therefore, we do not show data in the linear viscoelastic regime due to a lack of data for comparison against all fluid samples taken. However, available literature can provide useful information on the linear viscoelastic behaviour of polymer solutions. On the one hand, flexible polymer solutions mostly show $G' < G''$ across the entirety of the frequency range \citep{pereira2013,mohammadtabar2020,mitishita2022}. On the other hand, rigid polymer solutions (XG) at higher concentrations (e.g. 2000 ppm) show behaviours similar to a viscoelastic liquid, with the storage modulus $G' > G''$ at high frequencies, and $G' < G''$ at low frequencies \citep{mitishita2022}. Increasing the XG concentration leads to a gel-like behaviour \citep{pereira2013}, with $G'$ approximately constant with the frequency. The fact that the fluids share different rheological behaviours (both during SAOS and also extensional rheometry \citep{jaafar2010,pereira2013, warwaruk2021}), but have the same outcome in $\%DR$ shows that it does not seem possible to correlate DR to a rheological property of a given fluid \citep{mohammadtabar2020, warwaruk2021}.

\subsection{Experiment procedure} \label{S2-3}

We measure the turbulent velocity profile with LDA, the average pressure drop, flow rate and temperature ($T$) simultaneously. Initially, we circulate the fluid for approximately five minutes to ensure a steady-state. Afterwards, we acquire velocity data at 26 positions from $y/h \approx 0$ to the centreline at $y/h \approx 1$, so that the final velocity profile corresponds to one half of the duct height, assuming the velocity profile is symmetric. Depending on the data rate, at least $30000$ to $150000$ data points are acquired in the streamwise and wall-normal directions, respecitvely. The average data rate varied between $2500 Hz$ and $10000 Hz$. LDA measurements are performed at approximately 1.7, 2.2 and 2.7 $m/s$ for water, all concentrations of HPAM, XG at 150 and 500 ppm, and at 2.2, 2.7, 3.2 and 3.7 $m/s$ for XG 1500 ppm, totalling 22 velocity data sets. However, we do not show velocity and Reynolds stresses for all of them in the results section. Instead we focus on comparing data sets at similar Reynolds numbers. As mentioned previously, one polymer mix is used to perform all data sets at a fixed concentration. Another focus of this paper is the spectral analysis of  turbulent velocity fluctuations, which benefits from very high data rates. For this reason, we perform experiments only in non-coincidence mode, which provides at least double the data rate than coincidence mode. Furthermore, excluding the coincidence mode from the experimental plan reduces the data collection time, thus avoiding polymer degradation. The limitation of this method is that we are unable to present Reynolds shear stress measurements, which require data collection of $U$ and $V$ in coincidence mode.

To calculate turbulent drag reduction, we require the mean wall shear stress of the duct flow: 
\begin{equation}  \label{eq-tauw}
	\tau_w = \frac{\Delta P D_h}{4 L},
\end{equation}
where $\Delta P$ is the pressure drop measurement and $L$ is the length between each pressure tap. The wall shear stress is then used to calculate the percentage of drag reduction:
\begin{equation}  \label{eq-DR}
	\%DR = \frac{ \tau_{w,N} - \tau_{w,DR} }{ \tau_{w,N} } \times 100,
\end{equation}
where $\tau_{w,N}$ is the mean wall shear stress of the Newtonian solvent and $\tau_{w,DR}$ is the mean wall shear stress of the drag-reducing fluid, both calculated at approximately the same average flow rate. 

The density of all solutions is taken to be the same as water ($\rho$ = 999 \textit{kg/m$^3$}). The friction velocity is defined as $u_{\tau} = \sqrt{ \tau_w/ \rho }$. With measurements of $U_b$ and the friction velocity, we proceed to calculate the generalized Reynolds number and the frictional Reynolds number, respectively:
\begin{equation}  \label{eq-ReG}
	Re_G = \frac{\rho U_b D_h}{\eta_w},
\end{equation}
\begin{equation}  \label{eq-Retau}
	Re_\tau = \frac{\rho u_\tau h}{\eta_w},
\end{equation}
where $\eta_w = \tau_{w}/\dot{\gamma}_w$ is the viscosity of the fluid at the wall, and $\dot{\gamma}_w$ is the mean shear rate at the wall. We use the value of $\eta_w$ to calculate the (generalized) Reynolds number, similar to previous studies with shear-thinning fluids \citep{escudier2009a,owolabi2017,singh2017b,mitishita2021,mitishita2022}, due to the fact that the viscosity is not constant across the duct cross-section. The value of $\dot{\gamma}_w$ is computed with the mean wall shear stress $\tau_{w}$ and the constitutive equation fitted to the steady flow curve from the rheometer. After obtaining the values of $\dot{\gamma}_w$ and $\tau_{w}$, we can calculate $\eta_w$. Note that $Re_G$ = $Re$ for the water experiments, in which case $\eta_w$ is constant. The viscosity at the wall is used to define the wall unit $y^+_0 = \eta_w /\rho u_{\tau}$, with which we normalize the wall-normal coordinate $y$ as $y^+ = y/y^+_0$. \hyperref[table2]{Table~\ref*{table2}} shows the main experimental parameters for the duct flow of water, HPAM and XG solutions that are discussed in \hyperref[S3]{Section~\ref*{S3}}.

\begin{table*}[t]
		\centering
		\begin{tabular}{ *{8}{c} }
			\hline
			Concentration & $U_b$ [$m/s$] & $Re,Re_G$ [-] & $u_\tau$ [$m/s$] & $T$ [$^\circ$C]  &  $\tau_w$ [$Pa$] & $\eta_w$ [$Pa.s$]  & $DR$ [\%]  \\
			\hline
			\multicolumn{8}{c}{Water} \\
			\hline	
			- & 1.75 & $59 \times 10^3$  & 0.087 & 19.9 & 7.54 & $\approx 0.0009$ & - \\
			- & 2.26 & $77 \times 10^3$ & 0.109 & 20.2 & 11.80 & $\approx 0.0009$ & - \\
			- & 2.77 & $95 \times 10^3$ & 0.130 & 20.7 & 16.93 & $\approx 0.0009$ & - \\
			- & 3.79 & $132 \times 10^3$ & 0.171 & 21.3 & 29.28 & $\approx 0.0009$ & - \\
			\hline
			\multicolumn{8}{c}{HPAM} \\
			\hline	
			7.5 ppm & 1.86 & $65 \times 10^3$ & 0.078 & 19.0 & 6.07 & 0.00097 & 20 \\
			30 ppm & 1.84 & $64 \times 10^3$ & 0.061 & 21.6 & 3.69 & 0.00096 & 51 \\
			100 ppm & 1.72 & $43 \times 10^3$ & 0.050 & 22.5 & 2.45 & 0.0013 & 67 \\
			\hline
			\multicolumn{8}{c}{XG} \\	
			\hline
			150 ppm & 1.86 & $57 \times 10^3$ & 0.080 & 22.1 & 6.35 & 0.0011 & 16 \\
			500 ppm & 2.75 & $59 \times 10^3$ & 0.090 & 22.8 & 8.06 & 0.0016 & 52 \\
			1500 ppm & 3.71 & $42 \times 10^3$ & 0.108 & 23.3 & 11.7 & 0.0030 & 60 \\
			\hline
		\end{tabular} \caption{\label{table2} Experimental parameters for turbulent flow experiments of water, HPAM and XG. Water statistics and spectra are only shown for the $U_b=1.75$ $m/s$ data set. The remaining water data are presented as friction factor against $Re$, and are used for \%DR calculations for the polymer flows.}
\end{table*}

%%%%%%%%%%%%%%%%%%%%%%%%%%%%%%%%%%%%%%%%%%%%%%%%%%%%%%%%%%%%%%%%%%%%%%%%%%%%%%%%%%%%%%%%%%%%%%%%%%%%%%%%%%%%%%%%%%
\section{Results} \label{S3}

\subsection{Velocity and Reynolds stress profiles} \label{S3-1}

\begin{figure*}
	\centering
	\includegraphics*[width=\textwidth]{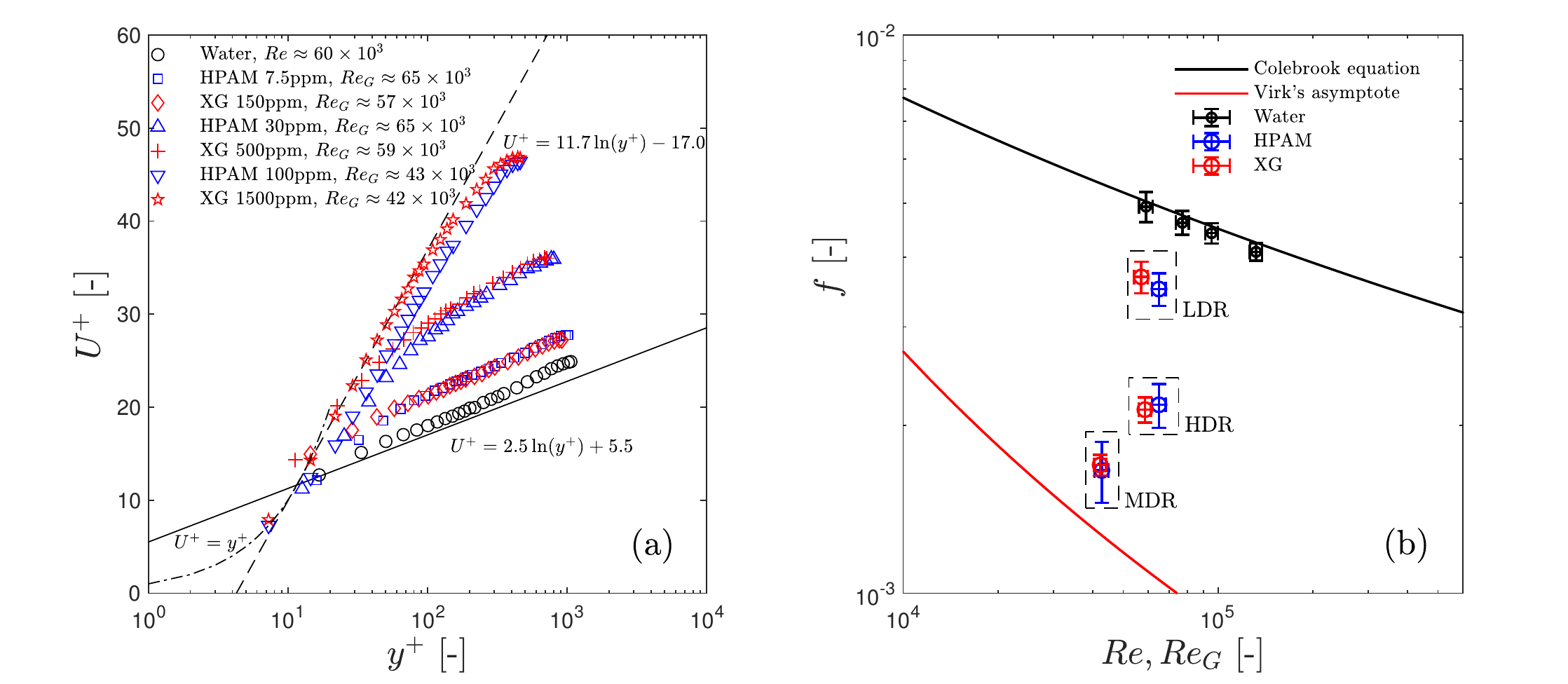}
	\caption{(a) Velocity profiles of water, HPAM and XG solutions. (b) Friction factor as a function of $Re_G$ (or $Re$ for water). Dot-dash lines represent the viscous sub-layer velocity profile, solid lines represent the logarithmic layer velocity profile for water, and the dashed lines show the Virk's MDR velocity profile.}
	\label{fig3}
\end{figure*}

Our instantaneous velocity data from 2-component LDA experiments are decomposed into a time-averaged and fluctuating component for analysis: the streamwise velocity component $U$ is given by the sum of the mean streamwise velocity $\langle U \rangle$ and the fluctuations $u$: $U = \langle U \rangle + u$. The same decomposition is applied to the wall normal velocity component: $V = \langle V \rangle + v$. We begin the results section with time-averaged velocity profiles normalized with the friction velocity ($U^+ = \langle U \rangle/ u_{\tau}$), plotted against $y^+$, and Fanning friction factor $f$ data of water and polymer solutions in \hyperref[fig3]{Figure~\ref*{fig3}}. We calculate $f$ with the wall shear stress and bulk velocity:
\begin{equation} \label{eq-f}
	f = \frac{2 \tau_w}{\rho U_b^2}.
\end{equation}
Futhermore, we present the Colebrook equation for Newtonian fluid and Virk's equation as an empirical approximation for the MDR limit:
\begin{equation} \label{eq-colebrook}
	\frac{1}{ \sqrt{f} }  = -4.0 \, \log_{10} \, \left( \frac{\epsilon/D_h}{3.7} + \frac{ 1.255 }{ Re_G \sqrt{f} } \right),
\end{equation}
\begin{equation} \label{eq-virk}
	\frac{1}{ \sqrt{f} }  = 19.0 \, \log_{10} \, \left( Re_G \, \sqrt{f} \right)  - 32.4.
\end{equation}
where the roughness $\epsilon$ is neglected for the test section, made of smooth acrylic. Information from the data sets shown in \hyperref[S3-1]{Section~\ref*{S3-1}} and \hyperref[S3-2]{Section~\ref*{S3-2}} are presented in table. 

\hyperref[fig3]{Figure~\ref*{fig3}}(a) depicts velocity profiles of water ($U_b = 1.75$ $m/s$), and 150 ppm XG and 7.5 ppm HPAM solutions at LDR, 500 ppm XG and 30 ppm HPAM solutions at HDR and 1500 ppm XG and 100 ppm HPAM solutions at MDR. To the best of our ability, we keep the generalized Reynolds numbers as close as possible to ensure a suitable comparison. \hyperref[fig3]{Figure~\ref*{fig3}}(b) shows the friction factor data against $Re_G$, and except from the MDR data which was obtained at a somewhat lower $Re_G$, the Newtonian, LDR and HDR datasets were acquired at similar Reynolds numbers. 

The error bars represent the propagation of uncertainties from the pressure drop, flow meter, duct dimensions and the rheometer measurements. Bear in mind that the definition of $Re_G$ for shear-thinning fluids takes into account the viscosity at the wall, and not the variation of viscosity throughout the cross-section of the duct. Furthermore, even though we estimate an uncertainty of 3\% for our rheometer measurements \citep{mitishita2021,mitishita2022}, additional errors may be present due to possible extrapolation of the viscosity of the fluids at shear rates beyond our rheometer measurements. 

Our water data are quite close to the Colebrook equation, and provide an adequate verification of our pressure measurements. The friction factors were also nearly the same for each drag reduction case, as can also be seen from the velocity profiles, where we observe matching velocity profiles with both types of polymeric fluids at approximately the same friction factor. The MDR data in particular does not quite reach the asymptote of Virk, but is nevertheless close enough that we consider it as MDR. Thus, we argue that the experimental conditions are adequate for comparing statistics and power spectral densities. 

The velocity of \hyperref[fig3]{Figure~\ref*{fig3}}(a) shows a near logarithmic velocity profile for both the Newtonian and LDR $U^+$, which was also observed in previous experimental studies with flexible and rigid polymer solutions. The values of the $U^+$ of water agree well with the theoretical profile in the log layer for Newtonian fluids. The limitations in computing a non-uniform $\tau_w$ and $U^+$ as a consequence of using a 2:1 duct in our experimental setup have been thoroughly discussed in our previous work \citep{mitishita2021, mitishita2022}. 

The slope of $U^+$ at LDR appears similar to the water case, but with higher values of velocity. In contrast to LDR, the velocity profiles at HDR and MDR lose their logarithmic dependence \citep{escudier2009a} as can be seen via indicator function $\xi = y^+ {\text{d}U^+}/{\text{d}y^+}$ (not shown here due to noise resulting from differentiating the $U^+$ experimental data). This fact has been shown by other experiments \citep{elbing2013,white2018}, confirming that the $U^+$ profiles at HDR and MDR are not logarithmic, and that Virk's asymptote is merely an approximation. Therefore, a clear indication of the transition from LDR to HDR is the disappearance of the logarithmic dependence of the $U^+(y^+)$ profile beyond the buffer layer, which is seen in both flexible and rigid polymer solutions.

\begin{figure*}
	\centering
	\includegraphics*[width=\textwidth]{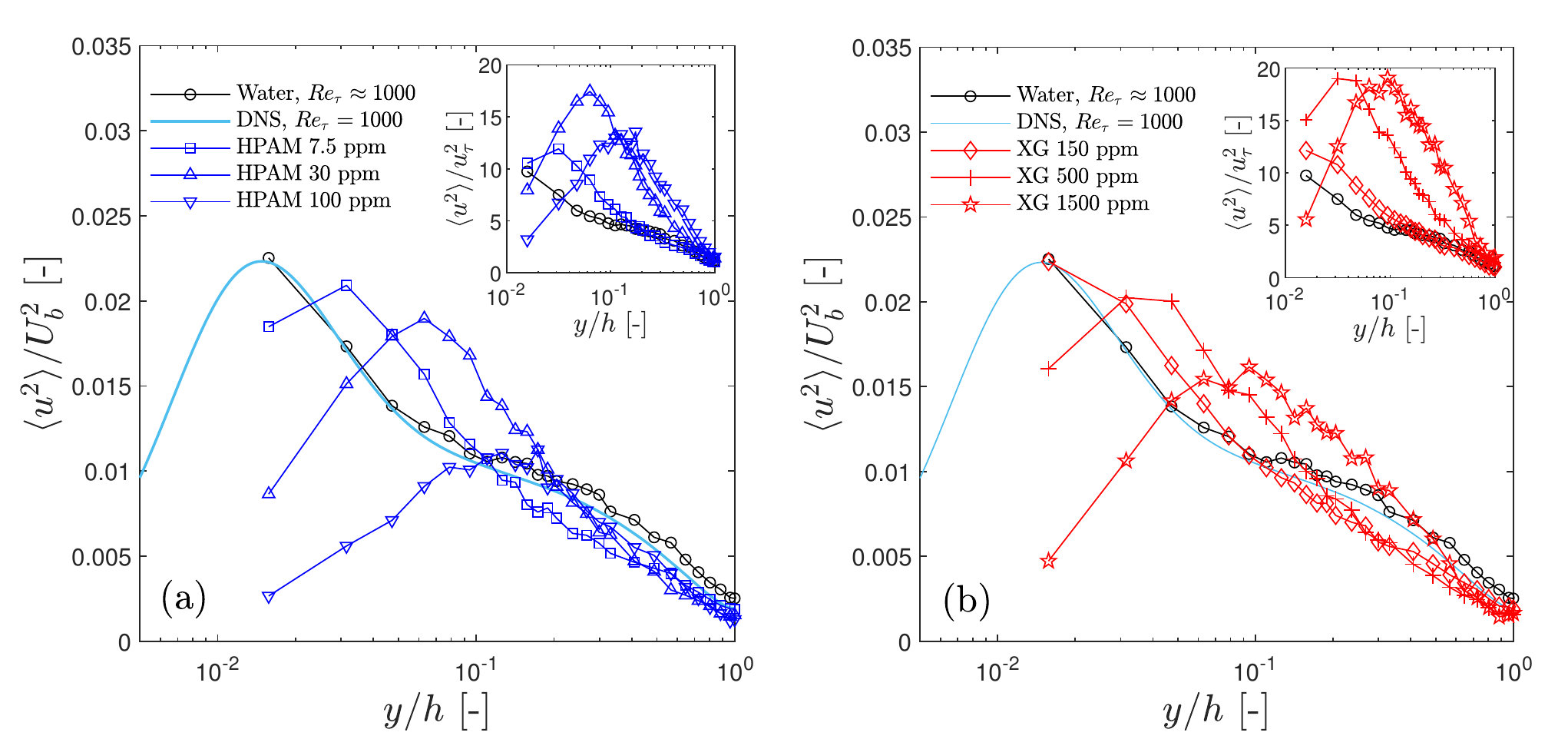}
	\caption{Streamwise Reynolds stress profiles of HPAM (a) and XG (b) solutions. Insets show the Reynolds stress profiles normalized by the friction velocity $u_{\tau}$}
	\label{fig4}
\end{figure*}

The streamwise Reynolds stresses normalized by the bulk velocity $\langle u^2 \rangle/U_{b}^2$ are shown in \hyperref[fig4]{Figure~\ref*{fig4}}(a) for HPAM solutions and \hyperref[fig4]{Figure~\ref*{fig4}}(b) for XG solutions, respectively. Also present for comparison are data points of water and Newtonian turbulent pipe flow DNS from \citet{elkhoury2013}. We see that at similar $Re_{\tau}$, the experimental data from water (taken at the mid-plane of the duct) accurately matches the pipe flow DNS data, even though the geometry is quite different, attesting to the accuracy of our experiments. 

The evolution of the velocity profile with the increase in concentration appear so be quite similar with both polymers. As the polymer concentration increases, the peak in $\langle u^2 \rangle/U_{b}^2$ decreases and moves away from the the wall, indicating a thickened buffer layer, a pattern also observed in the experiments \citep{shaban2018} with lower polyacrylamide concentrations, when normalized by the correspondent friction velocity of the solvent $u_{\tau0}$. In both situations, the polymers appear to absorb energy from log-layer and core regions as they stretch, and release that energy to near-wall eddies, a mechanism which has been verified by DNS of drag-reducing flows as well \citep{pereira2017b}.

The transitions from LDR to HDR, and from HDR to MDR are readily observed when $\langle u^2 \rangle$ is normalized by $u_{\tau}^2$, as reported in experimental \citep{warholic1999influence} and numerical investigations \citep{dallas2010,zhu2018}. Our $\langle u^2 \rangle/u_{\tau}^2$ data is presented in the insets of \hyperref[fig4]{Figure~\ref*{fig4}}. In both XG and HPAM solutions, the values of $\langle u^2 \rangle/u_{\tau}^2$ are nearly the same as the Newtonian counterparts in the log layer at LDR, a feature also seen in DNS from \citet{dallas2010}. As drag reduction increases from LDR to HDR, the peak in $\langle u^2 \rangle/u_{\tau}^2$ increases up to a maximum which is reached near the transition to HDR \citep{warholic1999influence}. The values of $\langle u^2 \rangle/u_{\tau}^2$ are larger than water in most of the turbulent boundary layer at HDR with both types of polymers. Beyond HDR, the peak in $\langle u^2 \rangle/u_{\tau}^2$ decreases and moves away from the wall, until MDR is achieved \citep{dallas2010}. 
The same trend is seen in our experiments with HPAM from the inset of \hyperref[fig4]{Figure~\ref*{fig4}}(a) as we increase the concentration from 7.5 ppm to 100 ppm. The XG solution appears to deviate from the trend as we move from HDR to MDR, where the peak in $\langle u^2 \rangle/u_{\tau}^2$ with the 1500ppm XG also moves further away from the wall but does not decrease when compared to the 500 ppm XG solution.

\begin{figure*}
	\centering
	\includegraphics*[width=\textwidth]{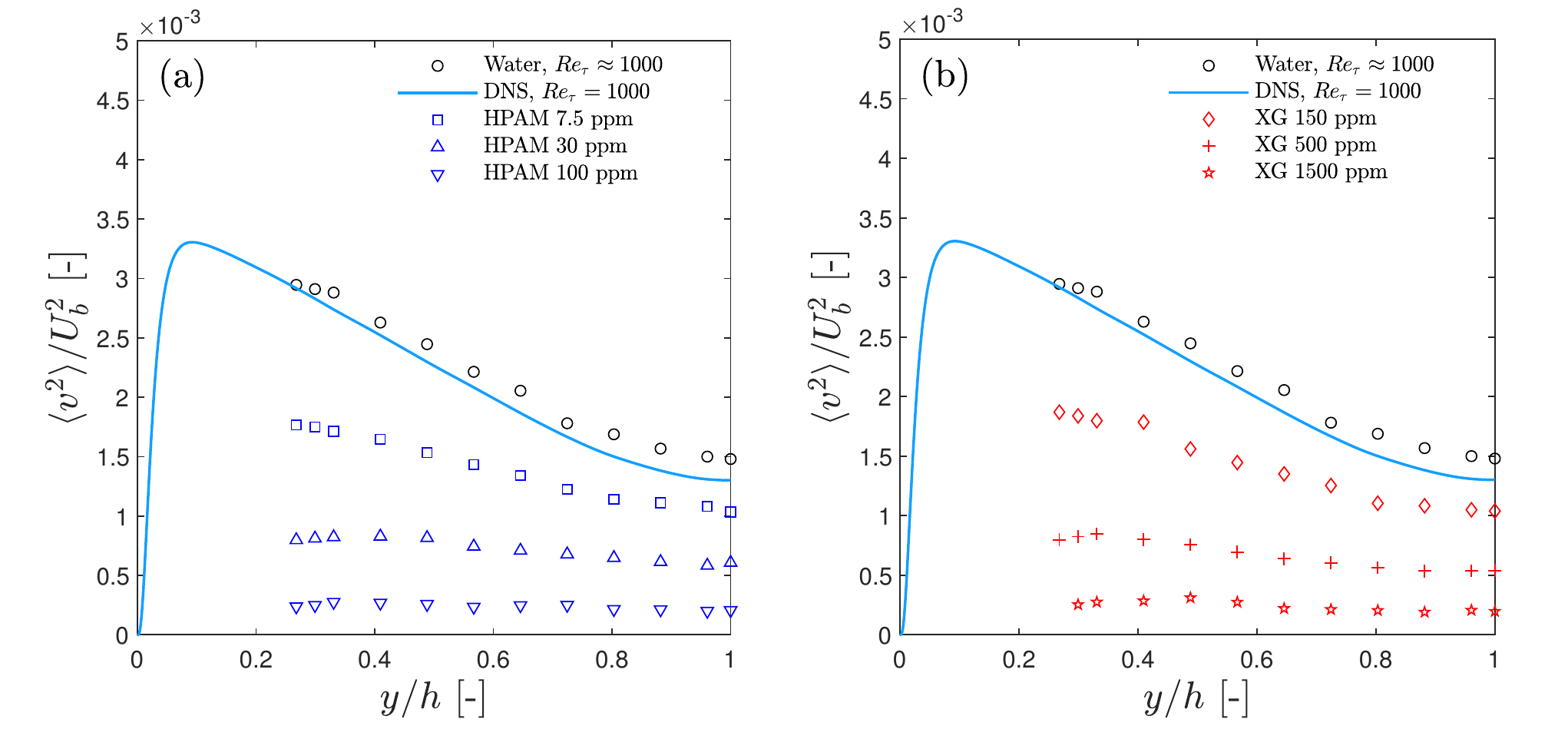}
	\caption{Wall-normal Reynolds stress profiles of HPAM (a) and XG (b) solutions.}
	\label{fig5}
\end{figure*}

\begin{figure*}
	\centering
	\includegraphics*[width=85mm]{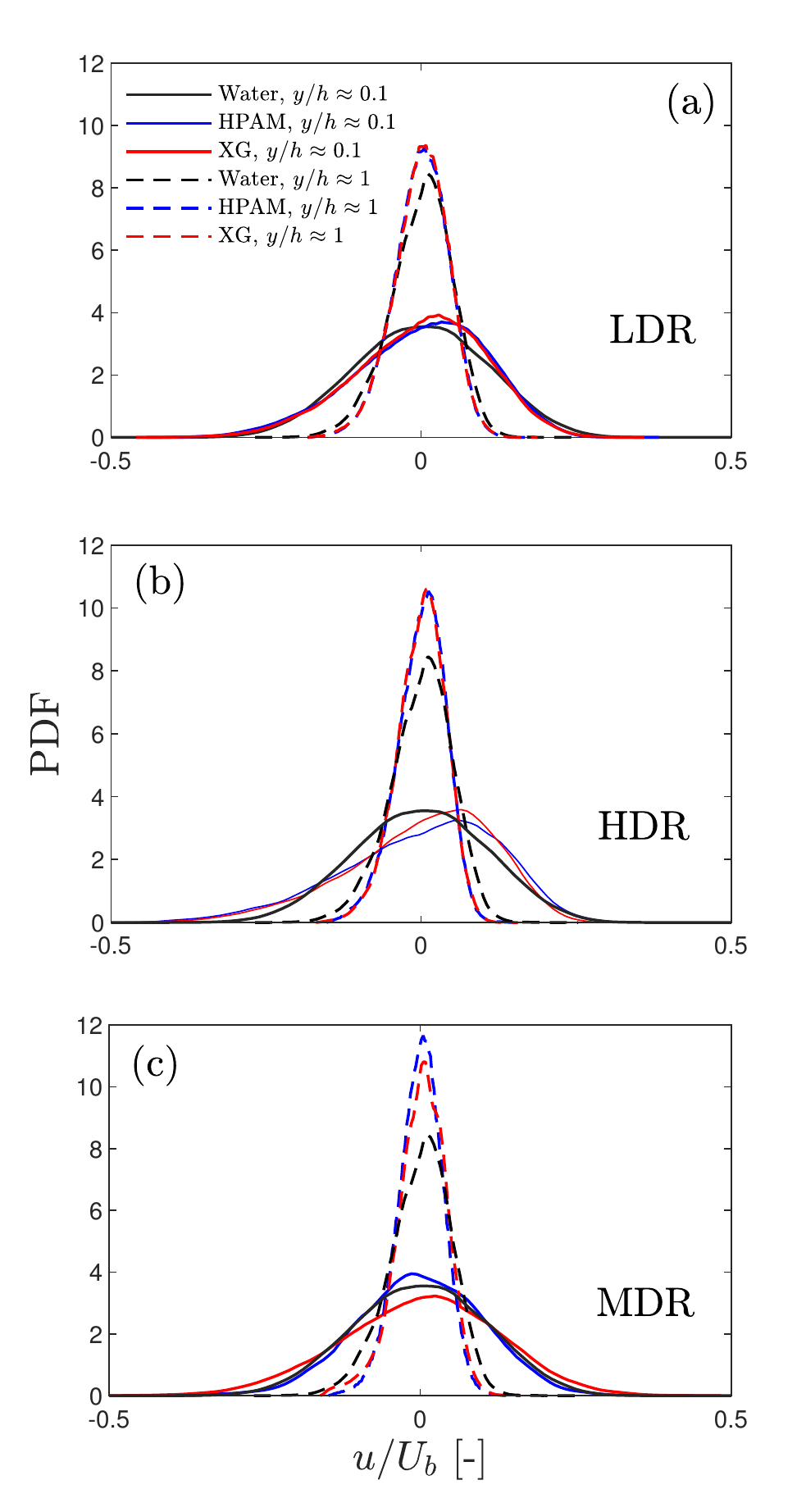}
	\caption{Probability density functions of water, HPAM and XG solutions at LDR (a), HDR (b) and MDR (c) at the position $y/h \approx 0.1$ and $y/h \approx 1$.}
	\label{fig6}
\end{figure*}

The wall-normal Reynolds stresses normalized by the bulk velocity $\langle v^2 \rangle/U_{b}^2$ are shown in \hyperref[fig5]{Figure~\ref*{fig5}}(a) for HPAM solutions and \hyperref[fig5]{Figure~\ref*{fig5}}(b) for XG solutions. The behaviour of the wall-normal Reynolds stresses agrees with previous studies, in which they suffer a significant decrease as the polymer concentration is raised. The MDR state is shown where $\langle v^2 \rangle/U_{b}^2$ are near zero, resulting in the Reynolds shear stresses $-\langle uv \rangle/U_{b}^2$ being near zero as well (not shown). In this state, turbulence is theorized to be mostly sustained by fluctuating elastic stresses from the polymers \citep{warholic1999}, which could explain why we do not see $\langle v^2 \rangle/U_{b}^2$ reach zero values. Overall, with the exception of the overall decrease in wall-normal fluctuations with DR, we do not see other hallmarks of the transition from each of the three DR states in the $\langle v^2 \rangle/U_{b}^2$, as we observe in both $U^+$ and $\langle u^2 \rangle/U_{b}^2$ data.

Our comparisons of velocity and Reynolds stress profiles show that the effects of the flexibility of polymer chains in turbulent drag reduction are more readily observed when the concentrations of the polymer solutions are relatively low. If we compare the 100 ppm HPAM and 150 ppm XG with similar shear viscosities, the HPAM solution achieves a much larger $\%DR$, even though the concentrations of XG is higher. This happens because HPAM is a flexible polymer, and viscoelastic effects in turbulent flows are significant even at low concentrations. Conversely, the effects of viscoelasticity are not very significant in XG solutions at low concentrations, as seen from multiple publications \citep{pereira2013,mohammadtabar2020,warwaruk2021}. Then, the drag reduction effect may be a consequence of its shear-thinning rheology \citep{singh2017b}. HDR and MDR are only possible with the XG solutions if we increase the concentration to near 1500 ppm. Therefore, viscoelastic effects with rigid polymer solutions appear to be significant only at high concentrations, with the consequence of increased shear viscosity. Non-negligible values of normal stress differences were measured with $0.2\%$ XG solutions \citep{escudier1999}, meaning that rigid polymer solutions can be viscoelastic, depending on the concentration. A plausible hypothesis for increased viscoelasticity of XG solutions in semi-dilute concentrations is the formation of aggregates, which is a polymer structure with composed of more than one polymer chain. The formation of XG aggregates may lead to a similar effect to an increase in molecular weight, increasing the viscoelasticity of the solution and thus enabling a higher $\%DR$ \citep{pereira2013, soares2020}.

Before we investigate the streamwise power spectral densities, we present the probability density functions (PDFs) of streamwise velocity fluctuations in \hyperref[fig6]{Figure~\ref*{fig6}}, in conditions of LDR (a), HDR (b) and MDR (c), for both HPAM and XG solutions. At LDR in \hyperref[fig6]{Figure~\ref*{fig6}}(a) and HDR in \hyperref[fig6]{Figure~\ref*{fig6}}(b), we observe that the probability distribution of $u$ of water does not change significantly by the addition of polymers. Near the wall, the distribution becomes somewhat skewed towards negative fluctuations, and the probability of positive values of $u$ becomes larger. In the centre of the duct, we observe a slight increase in the probability of near zero values of velocity fluctuations, which is expected of a drag-reducing flow. Nevertheless, there is almost zero difference between the distributions of XG and HPAM. At MDR in \hyperref[fig6]{Figure~\ref*{fig6}}(c) the picture is again mostly the same between both polymer solutions. The probability of near zero values of velocity fluctuations increases even further near the wall and at the centreline, which is in agreement with the decrease in the streamwise Reynolds stresses as observed in \hyperref[fig4]{Figure~\ref*{fig4}}. 

\subsection{Power spectral densities} \label{S3-2}

Streamwise power spectral densities (PSDs), $E_{uu}$, of velocity fluctuations $u(t)$ are presented in this section. We compare the PSDs of both XG and HPAM polymer solutions at LDR, HDR and MDR at three $y/h$ positions in the duct. For reference, we also show the PSDs of experiments with water at comparable Reynolds numbers. At a given position in the duct, the power spectral density is defined as an estimate of the energy distribution in frequency of velocity fluctuations, i.e.~$E_{uu} \propto |u_f(f)|^2$, where $u_f(f)$ is the Fourier transform of $u(t)$. We define the wavenumber space by $k_x = 2 \pi f / \langle U \rangle$ with the frequency data via Taylor's hypothesis \citep{pope2001}, which we consider valid if the root mean square of the velocity fluctuations $u_{rms}$ is less than 20\% of the mean local velocity $\langle U \rangle$ \citep{tran2010}. The wavenumber is then made dimensionless by multiplying by the duct half-height $h$. The PSD is made dimensionless dividing by $\nu_s U_b$, where $\nu_s$ is the kinematic viscosity of the solvent. 

We perform a linear interpolation of the non-uniform velocity-time signal from the LDA to obtain equally spaced data points. The interpolation frequency is the average data rate of the experiments \citep{toonder1997}. Furthermore, we truncate our results to 1/4 of the total wavenumber space to avoid filtering errors in the high wavenumber results \citep{ramond2000, mitishita2021}. For reference, we also present the PSD data of 500 ppm HPAM and 2000 ppm XG at MDR, originally published in our previous work \citep{mitishita2022}.

We start our analysis by presenting the streamwise PSDs at $y/h \approx 0.1$ in \hyperref[fig6]{Figure~\ref*{fig6}} in conditions of LDR (a), HDR (b) and MDR (c), for both HPAM and XG solutions. Solid yellow lines represent Kolmogorov's $k_x^{-5/3}$ scaling as a reference for the inertial range spectral scaling of Newtonian fluids. Deviations from the $k_x^{-5/3}$ are expected for the Reynolds numbers of the turbulent flow of water shown here \citep{smits2020}. We define the inertial range with \emph{approximate} limits of $5 \leq k_x \leq 25$, spanning approximately one-and-a-half decades in power density of streamwise velocity fluctuations of water. 

\begin{figure*}
	\centering
	\includegraphics*[width=90mm]{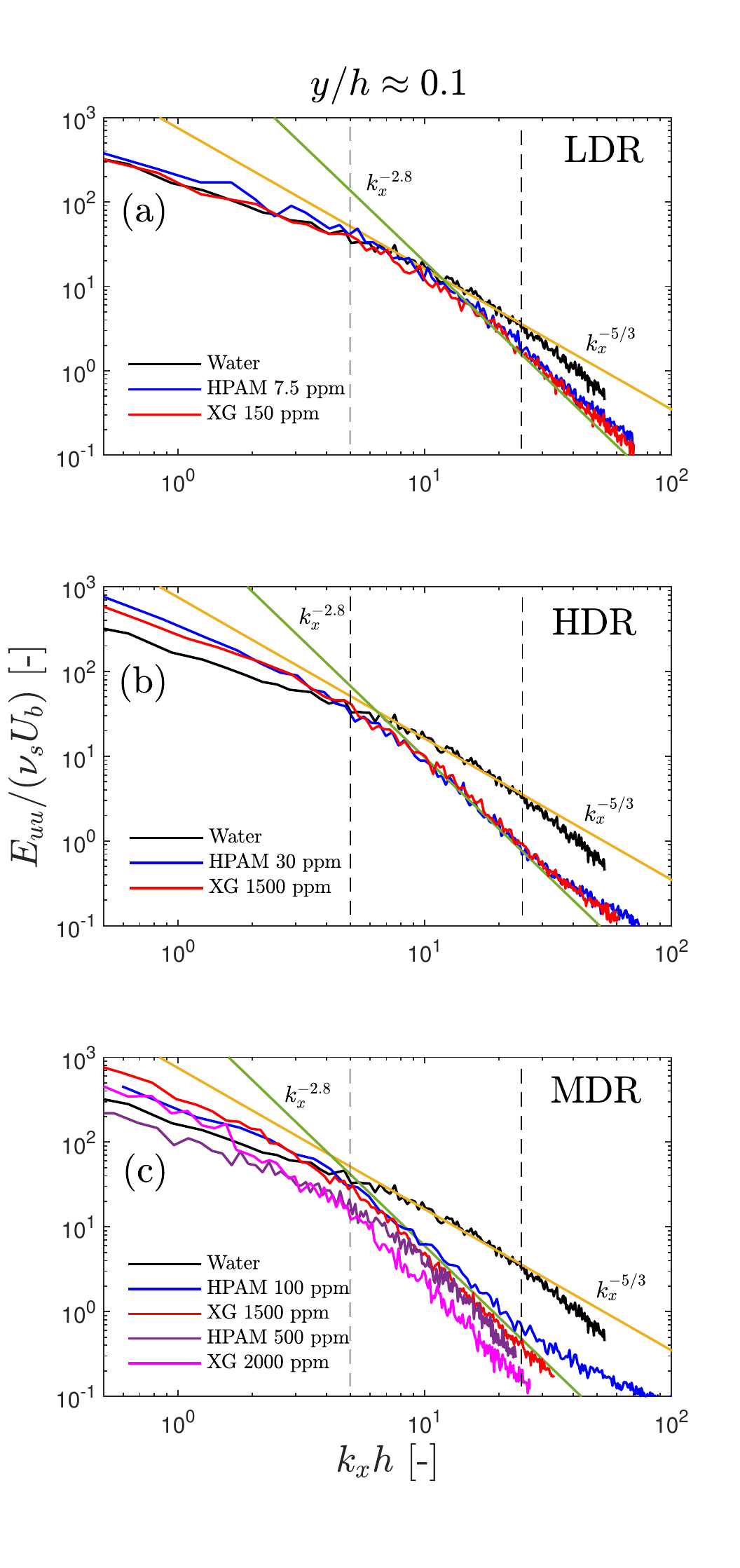}
	\caption{Streamwise power spectral densities of water, HPAM and XG solutions at LDR (a), HDR (b) and MDR (c) at the position $y/h \approx 0.1$. Dashed lines represent approximate limits of the inertial range of water PSDs, $5 \leq k_x \leq 25$. Data of 500 ppm HPAM and 2000 ppm is originally published in \citet{mitishita2022}}
	\label{fig7}
\end{figure*}

At LDR in \hyperref[fig7]{Figure~\ref*{fig7}}(a), there is a collapse of $E_{uu}$ values for HPAM and XG at all wavenumbers investigated. In comparison to the water spectrum in the inertial range, the HPAM and XG spectra decay in an approximate power law of $k_x^{-2.8}$ (green line) beyond $k_x h = 10$, indicating that the PSDs of small-scale turbulent structures (corresponding to large $k_x$ values) decrease slightly due to the action of the polymers in storing turbulent kinetic energy \citep{pereira2017b}. As we shall see, $k_x^{-2.8}$ fits well to the spectral decay with wavenumber of most PSD results in this paper, and is the focus of most of the discussion in \hyperref[S4]{Section~\ref*{S4}}. Interestingly, the transition from LDR to HDR in \hyperref[fig7]{Figure~\ref*{fig7}}(b) suggests that turbulent structures of length scales spanning the entirety of the inertial range are weakened by the polymers, with a good fit to the $k_x^{-2.8}$ power law. The PSDs of HPAM and XG at large $k_x$ decrease significantly when compared to the LDR state. Near the wall, the large-scale turbulent structures (corresponding to small $k_x$ values) appear to be mostly unaffected at HDR. 

At MDR, spectral decay due to polymer reaches wavenumbers lower than the inertial range limit, as shown in \hyperref[fig7]{Figure~\ref*{fig7}}(c). Furthermore, the $k_x^{-2.8}$ power law does not fit the HPAM data as well as the XG spectra, throughout most of the inertial range. The results of \hyperref[fig7]{Figure~\ref*{fig7}} hint that, at the same $\%DR$ and similar $Re_G$ values, no effects of the shear-thinning rheology on the near-wall streamwise power spectral densities can be seen, as we observe a collapse of $E_{uu}$ curves of both HPAM and XG in all three drag reduction levels probed. Even with the higher concentrations reported in \citet{mitishita2022}, the curves do not appear too far from each other. For insight on effects of shear-thinning rheology, we turn to the PSDs of turbulent fluctuations of Carbopol and water in \citet{mitishita2021}. The energy content in low wavenumbers in Carbopol is enhanced significantly, and even though a power law of $k_x^{-3.5}$ is observed at high wavenumbers, it can be argued that the Newtonian $k_x^{-5/3}$ scaling does not disappear, but is shifted to lower wavenumbers in comparison to water.

\begin{figure*}
	\centering
	\includegraphics*[width=90mm]{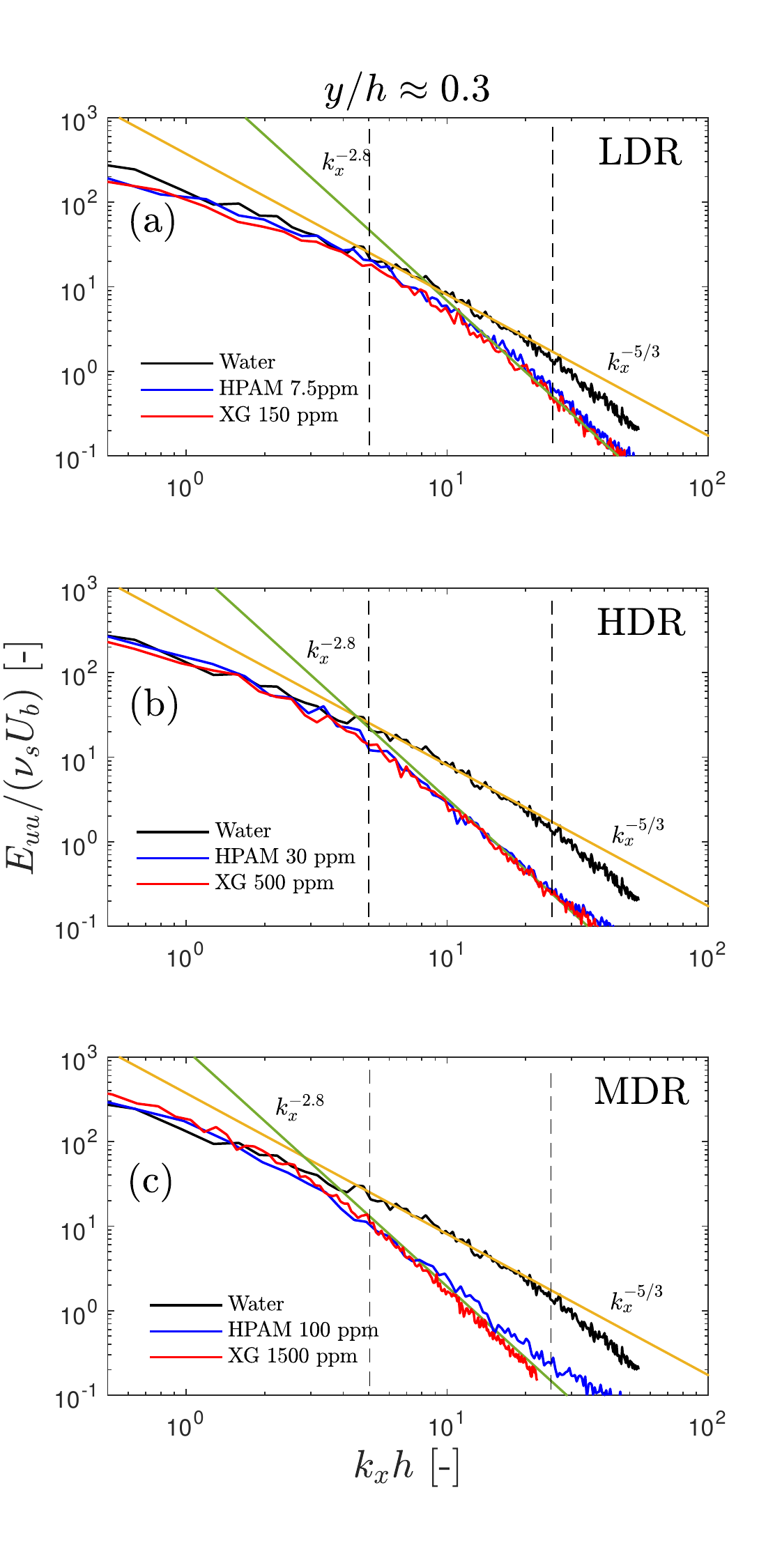}
	\caption{Streamwise power spectral densities of water, HPAM and XG solutions at LDR (a), HDR (b) and MDR (c) at the position $y/h \approx 0.3$. Dashed lines represent approximate limits of the inertial range of water PSDs, $5 \leq k_x \leq 25$.}
	\label{fig8}
\end{figure*}

\begin{figure*}
	\centering
	\includegraphics*[width=90mm]{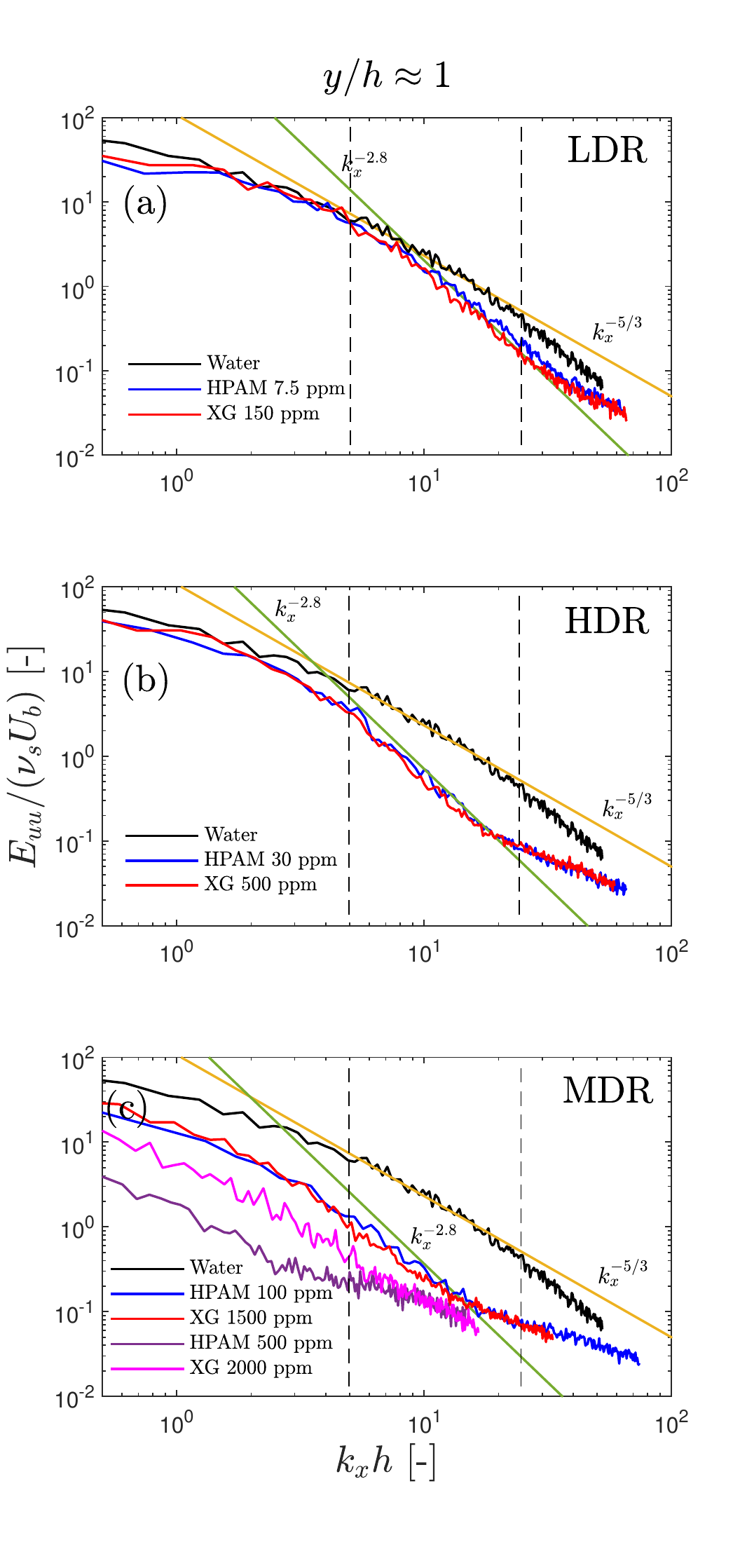}
	\caption{Streamwise power spectral densities of water, HPAM and XG solutions at LDR (a), HDR (b) and MDR (c) at the position $y/h \approx 1$. Dashed lines represent approximate limits of the inertial range of water PSDs, $5 \leq k_x \leq 25$. Data of 500 ppm HPAM and 2000 ppm is originally published in \citet{mitishita2022}}
	\label{fig9}
\end{figure*}

The streamwise PSDs in an intermediate position from the wall at $y/h \approx 0.3$ are presented in \hyperref[fig8]{Figure~\ref*{fig8}} at LDR (a), HDR (b) and MDR (c). The LDR results in \hyperref[fig8]{Figure~\ref*{fig8}}(a) are qualitatively similar to \hyperref[fig7]{Figure~\ref*{fig7}}, with the transition to LDR to HDR meaning that the entirety of the inertial range spectra scales with $k_x^{-2.8}$. As with the near wall spectra of \hyperref[fig7]{Figure~\ref*{fig7}}, there is also a significant decrease in power as we move from LDR to MDR. Low wavenumber structures are mostly affected at MDR only. The PSDs at the centreline of the duct are shown in \hyperref[fig9]{Figure~\ref*{fig9}} at LDR (a), HDR (b) and MDR (c), and the qualitative behaviour of the power spectra is again the same as \hyperref[fig7]{Figure~\ref*{fig7}} and \hyperref[fig8]{Figure~\ref*{fig8}}, in both LDR and HDR: we see a partial modification of the inertial range scaling of $k_x^{-5/3}$ to $k_x^{-2.8}$ at LDR, and at HDR the inertial range is fully modified by polymer additives. However, the MDR data deviates from the previous trends, as the $k_x^{-2.8}$ scaling does not fit well to the experimental results. Details of the fitting of the power-law exponents to the streamwise power spectra can be found in the Appendix.

The HPAM spectra lowers even further when we compare the PSDs of 100 ppm HPAM to the 500 ppm HPAM data from \citet{mitishita2022}, with most of the small-scale turbulent structures dampened by the polymer effects in the flow. Large-scale turbulent structures also lose a considerable amount of energy when the HPAM concentration is raised to 500ppm. Note that these differences are seen in a supposedly asymptotic state of MDR. If our results suggest that the differences in viscosity do not affect the PSDs of the polymer solutions, this implies increased elasticity effects with concentration. The concentration of the HPAM solutions used in our previous work is 5 times larger than the highest HPAM concentration in this paper \citep{mitishita2022}. We can speculate that the additional elasticity in the 500 ppm HPAM solution further dampens the streamwise velocity fluctuations in comparison to the 100ppm case. However, the overall drag does not decrease further due to increased polymer stress with concentration, which aid in sustaining a turbulent flow \citep{warholic1999}. The XG results from the present paper (1500 ppm) and from \citet{mitishita2022} (2000 ppm) do not vary as much when comparing the 100 ppm and 500 ppm HPAM, likely due to the fact that the XG concentration is raised by only 1.33 times.

%%%%%%%%%%%%%%%%%%%%%%%%%%%%%%%%%%%%%%%%%%%%%%%%%%%%%%%%%%%%%%%%%%%%%%%%%%%%%%%%%%%%%%%%%%%%%%%%%%%%%%%
\section{Discussion and conclusion} \label{S4}

We presented comparative time-averaged velocity profiles, Reynolds stress profiles and probability density functions of XG and HPAM solutions at the same level of drag reduction and Reynolds numbers. A high resolution 2-component LDA system was used to acquire the velocity signals over time at a large data rate suitable for spectral analysis. 

Experimental data presented in \hyperref[S3-1]{Section~\ref*{S3-1}} pointed out for $U^+$, $\langle u^2 \rangle/U_{b}^2$ and $\langle v^2 \rangle/U_{b}^2$, that drag reduction dynamics of both flexible (HPAM) and rigid (XG) polymer solutions show more similarities than differences. When comparing to the recent experimental study by \citet{warwaruk2021}, our results are in qualitative agreement with XG data at LDR and HDR. 

The $U^+$ data for both XG and HPAM show consistent results with literature, with the the velocity profile deviating from a logarithmic scaling beyond LDR, and approaching Virk's asymptote at MDR. From \hyperref[fig4]{Figure~\ref*{fig4}}, the $\langle u^2 \rangle/u_{\tau}^2$ profiles in HDR shows larger values than water throughout near the entirety of the buffer layer. Further, the only evident difference from XG and HPAM is the transition from HDR to MDR, where the peak in $\langle u^2 \rangle/u_{\tau}^2$ is observed to decrease in HPAM, but not in XG. \hyperref[fig5]{Figure~\ref*{fig5}} also shows consistent $\langle v^2 \rangle/U_{b}^2$ data to other experiments with rigid \citep{mohammadtabar2017, warwaruk2021} and flexible \citep{escudier2009a, shaban2018, warwaruk2021} polymer solutions. We show streamwise PDFs where the distribution of velocity fluctuations is, for the most part, the same for both types of polymer solutions.

The main novelty in this work is the direct comparison of turbulent flows of rigid and flexible polymer solutions via power spectral densities with high temporal resolution. Analogous to the surfactant drag reduction study from \citet{mitishita2022}, the slope of the power spectrum of both of our HPAM and XG experiments show an exponent near $-3$ (specifically $k_x^{-2.8}$), at all three drag reduction levels. Particularly at HDR and MDR, the Newtonian $k_x^{-5/3}$ spectra completely disappears from the inertial range, and is replaced by a PSD that scales with $k_x^{-2.8}$.

The approximate $k_x^{-3}$ power spectrum scaling of velocity fluctuations has been observed in inertialess and also inertial, high-$Re$ turbulence of viscoelastic fluids. At $Re \ll 1$, turbulent-like velocity fluctuations were initially observed in measurements of flow with dilute, flexible polymer solutions between parallel plates by \citet{groisman2000}. The general consensus is that cuvillinear streamlines, such as the aforementioned flow in parallel disks, or in serpentine microchannels \citep{burghelea2004} lead to elastic instabilities in the polymer flow. In the absence of inertia, the only sources of instability are the additional polymer stresses that arise from the stretching of the molecules as $Wi > 1$, or the elastic forces become dominant over viscous forces \citep{steinberg2021}. Flow analysis of elastic turbulence showed a scaling of power spectral densities $\approx k_x^{-3.5}$ \citep{groisman2000}.

The EIT phenomenon in flexible polymer solutions at high $Re$ has raised the possibility of similarities between inertialess and inertial, viscoelastic turbulent dynamics \citep{samanta2013}. Indeed, the $k_x^{-3}$ spectral scaling was also observed in DNS of turbulent flows \citep{valente2016}, DR experiments in grid turbulence \citep{vonlanthen2013}, and more recently in turbulent polymer jets \citep{yamani2021}. In those studies, the $k_x^{-5/3}$ inertial range scaling is replaced by $k_x^{-3}$, which was associated with the time-averaged strain rates which dominate the flow via dimensional arguments. Some channel flow DNS of turbulent drag reduction differ somewhat from experiments \citep{dubief2013, thais2013, pereira2017c}, with a steeper slope of the power spectrum between $-4$ and $-5$. Therefore, interaction between inertial and elastic instabilities result in PSDs that are alike to purely elastic, low $Re$ turbulence. 

Considering the spectra reported in studies of high-$Re$ EIT flows, the $k_x^{-2.8}$ slope of the PSDs of our duct flow experiments suggests the occurrence of EIT-like turbulent structures. Further, our results show that an approximate $k_x^{-3}$ scaling is common to turbulent drag reduction flows with both rigid and flexible polymer solutions. As drag reduction increases, the power spectra of larger length scales are modified by the polymers. The data suggests that the amount of drag reduction dictates the eddy length scale at which there is drop off in power when compared to water. The approximate transition from LDR to HDR happens when most of the inertial range is modified by the action of the polymers. Moreover, from the results in \hyperref[S3-1]{Section~\ref*{S3-1}} (especially \hyperref[fig6]{Figure~\ref*{fig6}}) it was expected that the streamwise PSDs of XG and HPAM would be very similar to each other.

However, our experiments cannot conclusively point out that our $k_x^{-2.8}$ slope of the PSDs of HPAM and XG are a consequence of EIT-like instabilities. Firstly, \citet{yamani2021} characterized the transition from laminar flows to EIT via visualization of the turbulent jets, where transitional turbulent structures are replaced with highly stretched structures as $Re$ increases. In comparison, \citet{samanta2013} increased $Re$ during pipe flow of a polymer solution, and noticed a sudden transition from the laminar state to a turbulent-like state, but at lower Reynolds numbers from those at which turbulence is observed in Newtonian fluids. We did not attempt a similar protocol, because of limitations in reaching the laminar regime in our setup. Secondly, it is unknown whether or not EIT can occur at the very high Reynolds numbers reported in this paper, or even with rigid polymer solutions such as XG \citep{xi2019}. For reference, the lowest $Re_G$ in the present paper is approximately $1.7 \times 10^4$ with XG, at which we also observe the $k_x^{-2.8}$ slope of the PSD (not shown). To obtain additional comparative data, experiments like those performed by \citet{samanta2013} and \citet{choueiri2018}, albeit with XG solutions, could be important future work. Thirdly, we do not have the possibility of visualizing flow structures the same way it was achieved with PIV by \citet{choueiri2018}. 

Our results indicate that EIT may arise during a turbulent flow of XG at the concentrations necessary for HDR and MDR. There are numerous reports of non-negligible viscoelasticity in XG at high concentrations, mostly from SAOS experiments in the linear viscoelastic regime \citep{pereira2013,soares2015,goyal2017,santos2020,mitishita2022}. In fact, our results show that differences between DR with rigid and flexible polymer solutions are more apparent at low concentrations. For example, when comparing the 100 ppm HPAM and 150 ppm XG results, inter-molecular interactions are not as frequent as with higher concentrations, and the flexibility of polymer chains can be correlated to higher DR. As we increase the XG concentrations, elastic effects may be enhanced from the formation intermolecular aggregates that may contribute to increase DR \citep{soares2020}.

The results here hint at the possibility that EIT-like turbulent structures may appear in turbulence with rigid polymer solutions, but only at high concentrations where elastic effects become more relevant. Additional rheometry and duct/pipe flow experiments are required to confirm EIT. Viscoelastic effects are evident at HDR and MDR, where the entirety of power spectra of the Newtonian inertial range is affected by the polymers. Thus, even though our experiments cannot quantitatively differentiate the DR mechanisms between XG and HPAM solutions, the overall dynamics of turbulence appear to be almost the same, from LDR to MDR.

%%%%%%%%%%%%%%%%%%%%%%%%%%%%%%%%%%%%%%%%%%%%%%%%%%%%%%%%%%%%%%%%%%%%%%%%%%%%%%%%%%%%%%%%%%%%%%%%%%%%%%%%%%%%%%%%%%
\section*{Acknowledgments} \label{Ackn}

This research was made possible by funding from Schlumberger and NSERC under the CRD program, project 505549-16. Experimental infrastructure was funded by the Canada Foundation for Innovation and the BC Knowledge Fund, grant number CFI JELF 36069. This funding is gratefully acknowledged. We kindly thank MI Swaco for supplying the HPAM polymer and CP Kelco for the donation of the xanthan gum used in this paper. R.S.M also acknowledges financial support from the University of British Columbia 4-Year Fellowship PhD scholarship program.

\section*{Conflict of interests} \label{int}

The authors report no conflict of interest.

\bibliography{Turbulent_viscoelastic_spectra_old}% Produces the bibliography via BibTeX.

\section*{Appendix: verification of experimental results} \label{App}

\begin{figure*}
	\centering
	\includegraphics*[width=90mm]{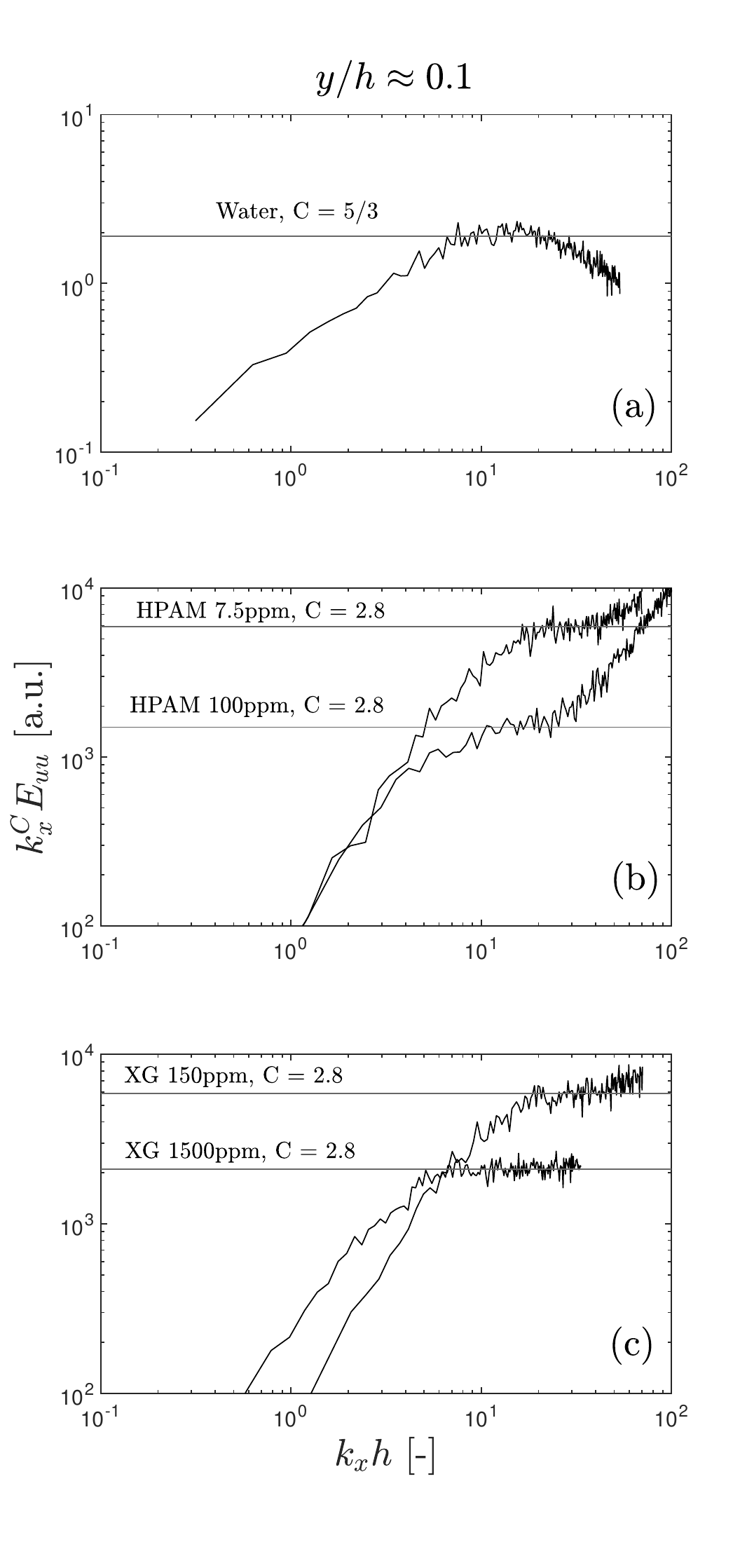}
	\caption{PSDs of velocity fluctuations, multiplied by $k_x^{C}$ for (a) water, (b) HPAM and (c) XG at $y/h \approx 0.1$. The constant C is the absolute value of slope of the power spectra of water (5/3) and polymer solutions (2.8).}
	\label{fig10}
\end{figure*}

We verify the accuracy of the power law fit to the streamwise PSDs ($E_{uu}$) of water, HPAM and XG by plotting $k_x^C E_{uu}$ (in arbitrary units, a.u.) against the dimensionless wavenumber $k_x h$ in \hyperref[fig10]{Figure~\ref*{fig10}}. A horizontal line indicates a good fit to the experimental data, depending on the value of the absolute value of the slope of the power spectrum C (5/3 for a Newtonian fluid and 2.8 for our polymer solutions). We do not present a verification of all PSD results. For the water PSD as shown in \hyperref[fig10]{Figure~\ref*{fig10}}, a good fit to Kolmogorov's $-5/3$ slope is found. For the drag-reducing fluids, the $-2.8$ slope is found to agree well with XG and HPAM results at LDR. At MDR (100 ppm HPAM), the fit is not as well representative of the computed $E_{uu}$ curve when compared to the 1500 ppm XG results.

\begin{figure*}
	\centering
	\includegraphics*[width=100mm]{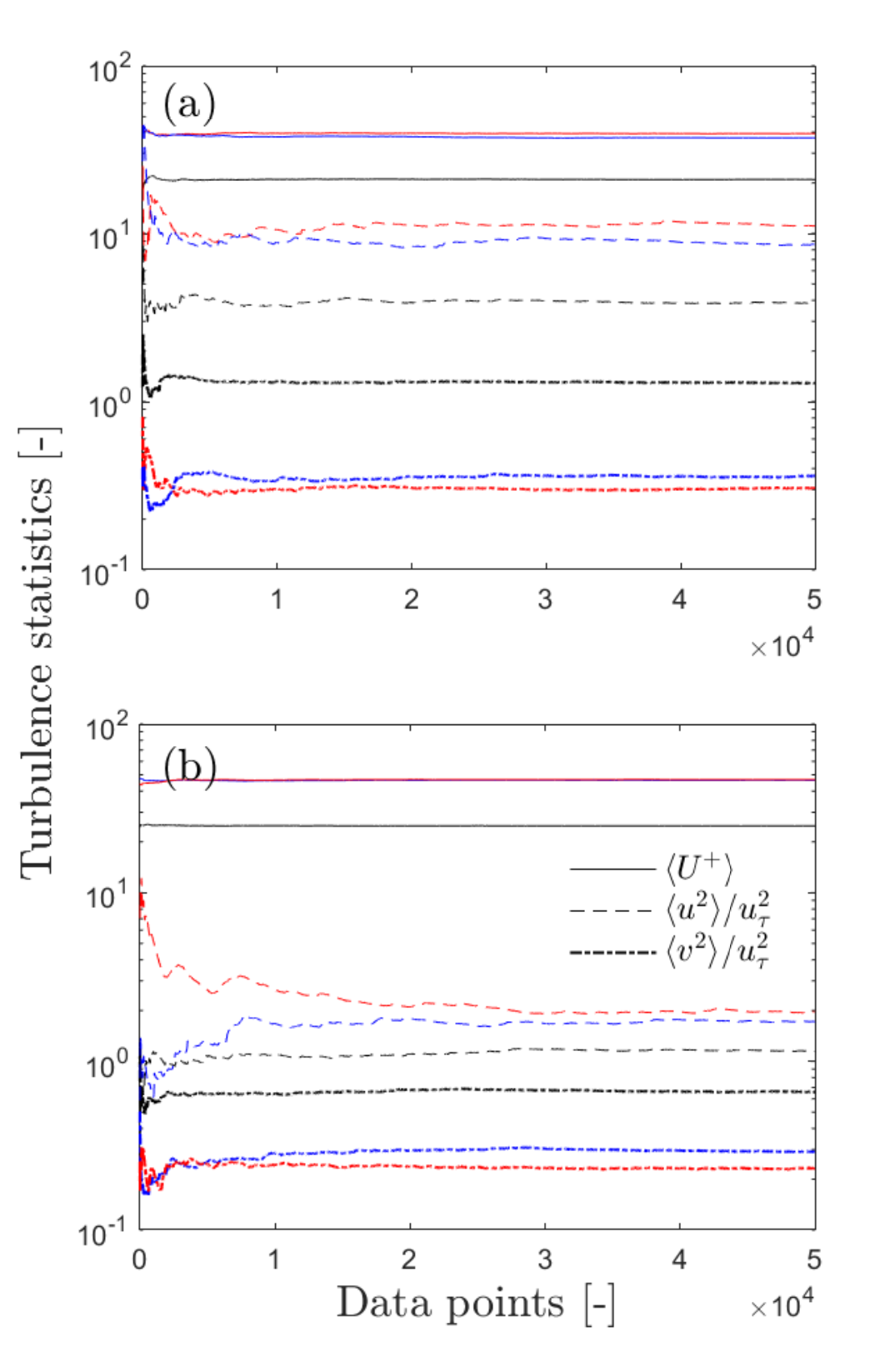}
	\caption{Convergence of turbulence statistics. Black lines represent water, blue lines represent 100 ppm HPAM and red lines represent 1500 XG solutions in position (a) $y/h \approx 0.3$ and (b) $y/h \approx 1$.}
	\label{fig11}
\end{figure*}

The convergence of turbulence statistics at $y/h \approx 0.3$ is examined in \hyperref[fig11]{Figure~\ref*{fig11}}. Only the water and MDR cases are shown here for brevity. We see that all turbulence statistics presented in this paper converge at approximately 30000 data points or less in both $y/h \approx 0.3$ and $y/h \approx 1$ positions.

\end{document}